\documentclass[12pt]{article}

\usepackage{amsmath}
\usepackage{amssymb}
\usepackage{geometry}
\usepackage{setspace}
\usepackage{hyperref}
\usepackage{graphicx}
\usepackage{booktabs}
\usepackage{natbib}

\newenvironment{tablenotes}
  {\begin{list}{}{\setlength{\leftmargin}{0pt}\setlength{\itemindent}{0pt}\setlength{\labelwidth}{0pt}\setlength{\labelsep}{0pt}\setlength{\topsep}{4pt}\footnotesize}}
  {\end{list}}

\title{Weighted Extensions of the Kolmogorov--Smirnov, Cram\'er--von Mises, and Anderson--Darling Tests for Assessing Covariate Balance}

{\small
\author{Ariel Linden, DrPH\\
University of California, San Francisco\\
Department of Medicine\\
Division of Clinical Informatics \& Digital Transformation (DoC-IT)\\
San Francisco, CA, USA\\
ariel.linden@ucsf.edu}
}

\date{}
\begin{document}

\maketitle

\section*{Abstract}
Assessing covariate balance is central to causal inference, but commonly used summary measures may miss distributional differences. I develop weighted extensions of the Kolmogorov--Smirnov (KS), Anderson--Darling (AD), and Cram\'er--von Mises (CVM) statistics and investigate their calibration when weights are externally supplied or estimated from treatment assignment. With exogenous weights, fixed-weight permutation produced approximately nominal Type I error. With estimated propensity-score weights, fixed-weight permutation was generally conservative, whereas re-estimating weights after unrestricted permutation could be substantially anticonservative when observed and resampled weighting regimes differed. Assignment-mechanism resampling was generally conservative under correct propensity-model specification but could become anticonservative under misspecification. Among the approaches examined for estimated propensity-score weights, fixed-weight permutation therefore provided the most reliable calibration, although it was generally conservative. AD retained strong sensitivity to tail discrepancies, while AD and CVM outperformed KS for diffuse discrepancies. For central discrepancies, KS was most powerful with exogenous weights, whereas AD was most powerful under estimated weighting. An inverse-probability-weighting example illustrates how these tests complement standardized mean differences and variance ratios. Weighted EDF tests provide complementary whole-distribution balance diagnostics. The methods are implemented in the Stata commands \texttt{kstest}, \texttt{adtest}, and \texttt{cvmtest}.

\section*{Keywords}

covariate balance; distributional tests; Kolmogorov--Smirnov test; Anderson--Darling test; Cram\'er--von Mises test; permutation inference; weighting

\section{Introduction}\label{sec1}

Covariate balance is a central diagnostic consideration in causal analyses that use propensity scores or related weighting methods to adjust for differences between treatment groups. Propensity-score weighting seeks to construct a weighted population in which treatment assignment is independent of measured baseline covariates, with the particular weighting scheme determining the target population and causal estimand \citep{rosenbaumrubin1983,robins2000,stuart2010}. Valid causal interpretation additionally requires assumptions such as conditional exchangeability, consistency, and adequate positivity or overlap; balance of measured covariates cannot establish these assumptions or rule out unmeasured confounding \citep{austin2015}. Nevertheless, inadequate balance after weighting indicates that the intended adjustment has not successfully removed observed differences between treatment groups, making balance assessment an essential diagnostic component of a weighting analysis. Examination of the resulting weight distribution and effective sample size provides complementary information about overlap and the information retained after weighting \citep{austin2015}.

Balance is most commonly assessed using descriptive measures such as standardized mean differences (SMDs), variance ratios, higher-order moments, and graphical comparisons of covariate distributions \citep{austin2009,austin2015,lindengraphic2015, lindensamuels2013, stuart2010}. These diagnostics answer different questions and should generally be considered jointly rather than as competing criteria. SMDs provide a scale-free summary of differences in means or proportions and have become a standard diagnostic in propensity-score analyses, while graphical methods can reveal discrepancies in spread, skewness, or other distributional features that moment-based summaries may miss. Recent methodological work has continued to develop alternative balance measures for weighted analyses, including diagnostics designed to accommodate different covariate types or to incorporate information about prognostic importance \citep{filla2022,kabata2024}. Thus, no single scalar diagnostic provides a complete characterization of covariate balance.

Empirical-distribution-function (EDF) measures offer a complementary way to assess balance because they compare entire covariate distributions. The Kolmogorov--Smirnov (KS) statistic, for example, has been used descriptively in propensity-score-weighted analyses to quantify the maximum discrepancy between weighted covariate distributions, alongside SMDs and graphical diagnostics \citep{austin2015,kabata2024}. More generally, three classical EDF goodness-of-fit statistics summarize distributional discrepancies in distinct ways: KS is determined by the largest pointwise difference between two empirical distributions \citep{kolmogorov1933,smirnov1948}; Cram\'er--von Mises (CVM) accumulates squared discrepancies across the distribution \citep{cramer1928,vonmises1931}; and Anderson--Darling (AD) additionally standardizes those discrepancies by their null variance, increasing sensitivity in the tails \citep{anderson1952,stephens1974}. These complementary sensitivities make the three statistics potentially useful for identifying forms of residual imbalance that may not be apparent from mean-based summaries alone. Their role is therefore not to replace SMDs or graphical assessment, but to provide additional information about whole-distribution balance.

Whether such discrepancies should also be accompanied by formal hypothesis tests is a separate question. Statistical significance tests for balance have long been criticized because their results depend on sample size and because the objective of balance assessment is often to characterize imbalance in the realized analytic sample rather than to test a superpopulation null hypothesis \citep{austin2009,austin2015}. At the same time, formal randomization- and permutation-based procedures for assessing covariate balance have an established methodological literature \citep{hansenbowers2008,gagnonbartshemtov2019}, and recent work continues to examine the statistical properties of balance criteria across different sample sizes and analytic settings \citep{hripcsak2025}. Formal distributional tests may therefore be useful as complementary diagnostics when their inferential target and limitations are stated clearly. In particular, a small \textit{p}-value should not by itself determine whether a weighting analysis has achieved acceptable balance, nor does a large \textit{p}-value establish causal validity.

A further inferential complication arises when the case weights are themselves estimated from treatment assignment and covariates. Defining a weighted distributional statistic for a supplied set of positive weights is distinct from establishing the appropriate reference distribution for that statistic when the weights are treatment-dependent. In observational studies, treatment assignments are not generally uniformly exchangeable, and randomization or permutation inference may require conditioning on, or explicitly representing, the treatment-assignment mechanism \citep{rosenbaum1984}. Related work has developed randomization-based inference under unequal treatment probabilities and methods that incorporate estimated propensity-score information into the assignment distribution \citep{bransonbind2019,pimentelhuang2024}. These considerations imply that holding estimated weights fixed while permuting treatment labels, re-estimating the weights after relabeling, and resampling treatment from an estimated assignment mechanism are different inferential procedures. The finite-sample consequences of these alternatives for weighted EDF balance tests do not appear to have yet been systematically characterized.

This study addresses both the construction and the inferential calibration of weighted distributional balance tests. First, I develop weighted versions of the KS, CVM, and AD statistics, each of which reduces to its classical counterpart when all case weights are equal. A common fixed-weight label-permutation procedure treats supplied case weights as realized attributes of the observations and provides inference conditional on those weights. Second, I investigate the operating characteristics of this procedure when weights are externally supplied versus estimated from treatment assignment, and compare it with two alternatives: re-estimation of propensity-score weights following label permutation and resampling from an estimated treatment-assignment mechanism. The simulation study evaluates Type I error and comparative sensitivity to central, tail, and diffuse distributional discrepancies under exogenously supplied and estimated treatment weights. It systematically examines weight variability, propensity-model specification, treatment-assignment strength, overlap and positivity, treatment prevalence, and the relationship between the effective-information regimes of the observed statistic and its resampling reference distribution. The study also evaluates whether the comparative sensitivities of KS, AD, and CVM differ between exogenous and estimated weighting. The proposed tests are illustrated as components of a broader covariate-balance assessment and are implemented in the Stata commands \texttt{kstest}, \texttt{adtest}, and \texttt{cvmtest} \citep{lindenadtest2025,lindencvmtest2025,lindenkstest2026}.

\section{Methods}\label{sec2}

\subsection{General framework}\label{sec2.1}

Consider two groups, indexed by $g\in\{1,0\}$, with sample sizes $n_1$ and $n_0$. In propensity-score applications, these groups typically represent treated and comparison subjects. Let $X_{gi}$ denote the value of a baseline covariate for observation $i$ in group $g$, and let $w_{gi}>0$ denote its supplied case weight. The objective is to assess whether the weighted distributions of $X$ are balanced between groups,
\begin{equation}
H_0: F_1^w = F_0^w,
\end{equation}
where $F_g^w$ denotes the weighted distribution of $X$ in group $g$.

Two distinct components of the testing problem are important. The first is statistic construction: given observed covariate values and a set of supplied positive weights, how should a distributional discrepancy between the two weighted groups be quantified? The weighted Kolmogorov--Smirnov (KS), Cram\'er--von Mises (CVM), and Anderson--Darling (AD) statistics developed below address this question. The second is null calibration: what reference distribution should be used to assess whether the observed value of that statistic is unusually large? These questions coincide in the conventional unweighted two-sample setting but need not do so when the weights are estimated using treatment assignment.

This distinction is particularly important in propensity-score-weighted analyses. A set of estimated weights is a function of the observed treatment assignments and covariates and therefore contains information about the treatment-assignment process. Treating those realized weights as fixed while reassigning treatment labels, re-estimating the weights after relabeling, and generating new treatment assignments from an estimated propensity-score model consequently define different resampling procedures. The weighted statistics themselves can be calculated under each procedure, but their inferential interpretations and operating characteristics do not need to be the same. The alternative resampling procedures considered in this study are described in Section~\ref{sec2.5} and evaluated empirically in Section~\ref{sec2.6}.

The tests considered here assess balance of one measured covariate at a time. They do not test the assumptions required for causal identification, establish the absence of unmeasured confounding, or determine whether the investigator has selected an appropriate adjustment set. They are intended as complementary balance diagnostics within a broader causal analysis.

\subsection{Weighted empirical distribution functions}\label{sec2.1b}

For group $g$, define the weighted empirical cumulative distribution function (ECDF) as
\begin{equation}
\hat F_g^w(x) = \frac{\sum_{i=1}^{n_g} w_{gi}\,\mathbb{1}(X_{gi}\leq x)}{\sum_{i=1}^{n_g} w_{gi}}.
\end{equation}
When $w_{gi}=1$ for every observation, this reduces to the ordinary empirical CDF,
\begin{equation}
\hat F_g(x) = \frac{1}{n_g}\sum_{i=1}^{n_g}\mathbb{1}(X_{gi}\leq x).
\end{equation}
For the AD statistic, define the weighted pooled ECDF as
\begin{equation}
\hat D^w(x) = \frac{\sum_{g\in\{1,0\}}\sum_{i=1}^{n_g} w_{gi}\mathbb{1}(X_{gi}\leq x)}{\sum_{g\in\{1,0\}}\sum_{i=1}^{n_g}w_{gi}}.
\end{equation}

Weights are required to be positive for observations included in the analysis. An observation assigned zero weight makes no contribution to the weighted ECDF and may equivalently be omitted. The weights need not be normalized to sum to a particular value. Because they enter the ECDFs through ratios, multiplication of all weights within a group by a common positive constant leaves the corresponding weighted ECDF unchanged.

\subsection{Weighted Kolmogorov--Smirnov and Cram\'er--von Mises statistics}\label{sec2.2}\label{sec2.3}

The ordinary two-sample KS statistic is the maximum absolute difference between the two empirical CDFs,
\begin{equation}
D = \max_x \left| \hat F_1(x)-\hat F_0(x) \right|.
\end{equation}
Its weighted analogue is obtained directly by substituting the weighted ECDFs,
\begin{equation}
D^w = \max_x \left| \hat F_1^w(x)-\hat F_0^w(x) \right|.
\end{equation}
Thus, $D^w$ measures the largest pointwise discrepancy between the two weighted covariate distributions. It reduces exactly to the ordinary KS statistic when all weights are equal.

Let $x_{(1)},\ldots,x_{(m)}$ denote the ordered distinct values in the pooled sample. The two-sample CVM statistic considered here accumulates squared discrepancies between the two ECDFs across their support,
\begin{equation}
CVM = \sum_{k=1}^{m-1} \left[ \hat F_1(x_{(k)}) - \hat F_0(x_{(k)}) \right]^2.
\end{equation}
The weighted statistic is
\begin{equation}
CVM^w = \sum_{k=1}^{m-1} \left[ \hat F_1^w(x_{(k)}) - \hat F_0^w(x_{(k)}) \right]^2.
\end{equation}
As with KS, the weighted CVM statistic therefore follows directly from replacing the ordinary ECDFs with their weighted counterparts. Neither statistic requires an additional variance-standardization term, and both reduce exactly to their unweighted forms when all weights are equal.

\subsection{Weighted Anderson--Darling statistic}\label{sec2.4}

The AD statistic differs from KS and CVM by standardizing the discrepancy between the group-specific ECDFs according to its position in the pooled distribution. Let $\tau_k$ denote the number of observations tied at $x_{(k)}$. In the unweighted setting,
\begin{equation}
AD = \sum_{k=1}^{m-1} \left[ \frac{ \hat F_1(x_{(k)}) - \hat F_0(x_{(k)}) }{ \sqrt{ \hat D(x_{(k)}) \{1-\hat D(x_{(k)})\}/(2N) } } \right]^2 \tau_k,
\end{equation}
where $N=n_1+n_0$.

Weighting requires an additional choice because $N$ no longer necessarily represents the effective information in the sample. I use the Kish effective sample size within each group,
\begin{equation}
n_{e,g} = \frac{ \left(\sum_{i=1}^{n_g}w_{gi}\right)^2 }{ \sum_{i=1}^{n_g}w_{gi}^2 }, \qquad n_e=n_{e,1}+n_{e,0},
\end{equation}
which equals $n_g$ when all weights within group $g$ are equal. The weighted AD statistic is then
\begin{equation}
AD^w = \sum_{k=1}^{m-1} \left[ \frac{ \hat F_1^w(x_{(k)}) - \hat F_0^w(x_{(k)}) }{ \sqrt{ \hat D^w(x_{(k)}) \{1-\hat D^w(x_{(k)})\}/(2n_e) } } \right]^2 \tau_k.
\end{equation}
The tie multiplier remains the raw number of observations sharing $x_{(k)}$, as in the corresponding unweighted discrete formulation. When all weights equal one, $n_e=N$ and the weighted statistic reduces to its unweighted form.

The use of Kish effective sample size above is a construction choice for the weighted AD statistic rather than a general variance estimator for a particular weighting procedure. It is intended to reflect the reduction in effective information associated with unequal weights while requiring no model for how those weights were generated. It should therefore be distinguished from the separate question of how the reference distribution of $AD^w$ is constructed. For any resampling procedure in which the weights vary across replicates, $n_e$ is recalculated from the weight vector used in that replicate.

\subsection{Resampling inference}\label{sec2.5}\label{sec:inference}

None of the three weighted statistics is compared with a tabulated asymptotic null distribution. Instead, \textit{p}-values are obtained from resampling reference distributions. Because the weights may either be externally supplied or estimated from treatment assignment, I distinguish three procedures.

\subsubsection{Fixed-weight label permutation}

The primary procedure implemented by the proposed tests treats the supplied case weights as fixed realized attributes of the observations. The observed statistic $T_{\mathrm{obs}}$, where $T$ denotes $D^w$, $CVM^w$, or $AD^w$, is calculated from the observed group labels, covariate values, and weights. In each of $R$ replicates, the group labels are randomly reassigned among observations while preserving the observed group sizes $n_1$ and $n_0$. Covariate values and case weights remain paired with their original observations. The statistic is recalculated after each relabeling, and the permutation \textit{p}-value is
\begin{equation}\label{eq:permp}
p = \frac{ 1+\#\{r:T_r^*\geq T_{\mathrm{obs}}\} }{ R+1 },
\end{equation}
where $T_r^*$ denotes the statistic from resampling replicate $r$.

This procedure defines inference conditional on the supplied weight vector. It does not re-estimate a weighting model or propagate uncertainty arising from estimation of that model. When weights are externally generated independently of group assignment, the procedure corresponds directly to permutation of labels among fixed covariate--weight pairs. When weights have instead been estimated from treatment assignment and covariates, the same algorithm remains well defined, but its reference distribution conditions on weights that were themselves estimated using the observed treatment assignments. Its finite-sample calibration in that setting is therefore an empirical question examined in the simulation study rather than assumed from the conventional exchangeability argument.

Fixed-weight label permutation is the procedure implemented by the Stata commands \texttt{kstest}, \texttt{adtest}, and \texttt{cvmtest}; the commands accept user-supplied case weights and do not estimate or re-estimate the weighting model.

\subsubsection{Weight re-estimation after label permutation}

A seemingly natural alternative when weights arise from a propensity-score model is to re-estimate that model after each permutation. Under this procedure, treatment labels are randomly reassigned while preserving group sizes; the propensity-score model is then refitted using the permuted labels, new weights are constructed from the refitted model, and the weighted statistic is recalculated.

This procedure differs fundamentally from fixed-weight permutation because the reference statistic is evaluated using a different weight vector in every replicate. Moreover, unrestricted permutation breaks the systematic treatment--covariate relationship represented in the observed data before the propensity-score model is refitted. Consequently, the resampled weight distribution need not resemble the observed weighting regime. The simulations below examine the resulting finite-sample calibration directly.

\subsubsection{Assignment-mechanism resampling}

A third approach represents treatment assignment explicitly. Let
\begin{equation}
e(X_i)=P(A_i=1\mid X_i)
\end{equation}
denote the propensity score and let $\widehat e(X_i)$ be its estimate from the observed data. In each resampling replicate, a new assignment is generated independently for each observation according to
\begin{equation}
A_i^* \sim \operatorname{Bernoulli}\{\widehat e(X_i)\}.
\end{equation}
The weights appropriate to $A_i^*$ are then constructed using the same fitted assignment probabilities $\widehat e(X_i)$; the propensity-score model is not refitted within the resampling replicate. The weighted statistic is recalculated. Unlike label permutation, this procedure does not preserve the observed treatment-group sizes.

This is an assignment-mechanism or conditional-randomization-type resampling procedure rather than a permutation test. Its interpretation depends on the fitted propensity-score model: if that model does not adequately represent the treatment-assignment mechanism, the resampling distribution may likewise be misspecified. The simulation study therefore evaluates this procedure under both correctly specified and misspecified propensity-score models.

The three procedures target related but distinct reference distributions. Fixed-weight permutation conditions on the realized weights; weight re-estimation asks what results when the weighting procedure is reapplied after unrestricted relabeling; and assignment-mechanism resampling constructs a reference distribution from an explicit model for treatment assignment. These procedures are not assumed to be inferentially equivalent. Their operating characteristics are compared empirically below.

\subsection{Covariate type and interpretation}\label{sec:covtype}

The proposed EDF statistics are most naturally applied to continuous covariates because their principal advantage is the ability to detect differences in distributional shape that may not be summarized adequately by differences in means. Ordered discrete covariates can also be analyzed; repeated values are accommodated by the empirical distribution functions, with the AD statistic additionally retaining its tie multiplier.

For binary covariates, the empirical distribution contains only one nontrivial difference. A weighted difference in proportions or corresponding standardized difference therefore provides a simpler and more interpretable balance diagnostic. When formal testing is desired, an appropriate weighted test of proportions or chi-square test may be used.

Unordered nominal covariates likewise are not naturally suited to EDF-based tests because their categories have no inherent ordering. Category-specific weighted proportions and appropriate standardized measures can be used for descriptive assessment, while a weighted chi-square test can provide an overall formal comparison across categories. Accordingly, the proposed EDF tests are most appropriate as complementary diagnostics for continuous and ordered covariates rather than as replacements for established balance measures across all covariate types.

\subsection{Simulation study}\label{sec2.6}

\subsubsection{Aims and overall design}

The simulation study was organized according to the ADEMP framework of \citet{morris2019}: aims, data-generating mechanisms, estimands or inferential targets, methods, and performance measures. Two complementary components were conducted. Simulation I evaluated the weighted KS, AD, and CVM statistics when case weights were generated independently of group assignment, assessing Type I error, comparative power against central, tail, and diffuse distributional discrepancies, and sensitivity to weight variability. Simulation II evaluated treatment-dependent inverse-probability weights and compared the three resampling procedures described in Section~\ref{sec:inference} under varying treatment-assignment strength, propensity-model specification, overlap, weight variability, and treatment prevalence.

The primary calibration target was Type I error at $\alpha=.05$. Definitive comparisons used 2,000 Monte Carlo replications, giving a Monte Carlo standard error of .0049 when the true rejection probability is .05 \citep{morris2019} and an approximate 95\% Monte Carlo interval of .040--.060. Unless otherwise stated, definitive comparisons used $R=500$ reference-distribution replicates. All simulations were conducted in Stata 19. Table 1 summarizes the design and purpose of the two simulation components.

\subsubsection{Exogenously supplied weights}

 For total sample size $n$, observations were allocated to Group 1 ($n_1=0.4n$) or Group 0 ($n_0=0.6n$). Each observation was assigned a quantile position $p_{gi}\sim U(0,1)$, and a continuous covariate was generated as
\begin{equation}
Y_{gi} = \mu+\sigma\Phi^{-1}(p_{gi}) + I(g=1)\delta(p_{gi}),
\end{equation}
where $\Phi^{-1}$ is the standard-normal quantile function, $\mu=50$, $\sigma=10$, and $\delta(\cdot)$ is a scenario-specific discrepancy function applied only to Group 1. Under the null, $\delta(p)\equiv0$. For centrally and tail-located discrepancies, $\delta(p)$ was a piecewise-linear function equal to $-\Delta$ at $p=p_L$, $+\Delta$ at $p=p_U$, and zero at the anchor points $\{0.0005,p_L-w,0.5,p_U+w,0.9995\}$, with linear interpolation between adjacent anchors. The central discrepancy used $p_L=.25$, $p_U=.75$, $w=.05$, and $\Delta=6$; the tail discrepancy used $p_L=.05$, $p_U=.95$, $w=.02$, and $\Delta=50$. The diffuse discrepancy used $\delta(p)\equiv1$, producing a uniform location shift.

Case weights were generated independently of $Y$, group assignment, and $p$:
\begin{equation}
w_{gi}=\exp(\eta_{gi}), \qquad \eta_{gi}\sim N\left(-\frac{1}{2}\sigma_\eta^2,\sigma_\eta^2\right),
\end{equation}
where
\begin{equation}
\sigma_\eta^2=\log(1+c^2), \qquad c^2=\frac{1}{r_e}-1,
\end{equation}
and $r_e=n_e/n$ denotes the target Kish effective-sample-size ratio. The principal simulations used $r_e=.60$. Because these weights are generated independently of group assignment, Simulation I isolates the behavior of the weighted statistics and fixed-weight permutation procedure without introducing treatment-dependent weight estimation.

Total sample size varied over $n\in\{1{,}000,2{,}000,3{,}000,4{,}000\}$, with 2,000 Monte Carlo replications and $R=1{,}000$ fixed-weight permutations per condition. The weighted KS, AD, and CVM tests were evaluated on the same simulated dataset within each replication, allowing their rejection outcomes to be compared on paired data. A sensitivity analysis fixed $n=2{,}000$ and repeated the four scenarios at $r_e\in\{.20,.40,.60,.80,.90\}$, assessing whether the comparative behavior of the three statistics and calibration of fixed-weight permutation depend materially on the degree of exogenous weight variability.

\subsubsection{Estimated treatment weights}

Simulation II generated a continuous baseline covariate $X_1\sim N(0,1)$ and assigned treatment according to
\begin{equation}
A_i\sim \operatorname{Bernoulli}\{e(X_{1i})\},
\end{equation}
with
\begin{equation}
\operatorname{logit}\{e(X_{1i})\} = \beta_0+ \lambda \left( 0.90X_{1i}+0.25X_{1i}^2 \right),
\end{equation}
where $\beta_0$ was selected to obtain the specified marginal treatment prevalence and $\lambda$ controlled the strength of the treatment--covariate relationship. Unless otherwise stated, treatment prevalence was approximately 0.50, $\lambda=1$, and $n=500$ throughout Simulation II.

\label{sec:sim2dgm}Inverse-probability-of-treatment weights targeting the average treatment effect were constructed from the estimated propensity score $\widehat e(X_{1i})$ as
\begin{equation}
w_i=\frac{A_i}{\widehat e(X_{1i})}+\frac{1-A_i}{1-\widehat e(X_{1i})},
\end{equation}
following the standard inverse-probability-weighting formulation \citep{robins2000}. Under correct specification, the fitted logistic propensity-score model included both $X_1$ and $X_1^2$; under misspecification, the quadratic term was omitted. This design permits the weighting model to be varied independently of the known treatment-assignment mechanism. Under the null calibration experiments, the outcome/covariate distribution being compared contained no treatment-induced discrepancy.

The three resampling procedures defined in Section~\ref{sec:inference}---fixed-weight label permutation, weight re-estimation after label permutation, and assignment-mechanism resampling---were compared. An exogenous-weight condition corresponding to Simulation I was included as a reference where appropriate.

\subsubsection{Calibration experiments}

Calibration was examined through complementary experiments designed to distinguish absolute weight variability from differences between the weighting regimes represented by the observed and reference statistics. Unless otherwise noted, these experiments used 1,000 Monte Carlo replications, $R=500$ resampling replicates, and $n=500$.

Treatment-assignment strength was varied over $\lambda\in\{0.5,1.0,1.5\}$ under correctly specified and misspecified propensity models, assessing how calibration changes as the treatment--covariate relationship strengthens and the observed weights become more variable. To examine observed--reference weight-regime mismatch more directly, treatment labels were also permuted within $S\in\{1,2,4,8,16\}$ quantile strata of $X_1$, progressively preserving the observed treatment--covariate relationship and creating intermediate reference weight distributions between unrestricted permutation and the observed-data weighting regime while holding the underlying treatment-assignment strength fixed. The mismatch was summarized using the Kish ratio $r_e=n_e/n$ as
\begin{equation}
\Delta r_e = \overline{r}_{e,\mathrm{resamp}} - r_{e,\mathrm{obs}},
\end{equation}
where $\overline{r}_{e,\mathrm{resamp}}$ is the mean Kish ratio across resampling replicates; positive values indicate that the reference distribution is generated under more nearly uniform weights, and hence a larger effective sample size, than the observed statistic. Restricted permutations were required to retain adequate representation of both treatment groups within each stratum: a minimum of 3 treated and 3 control units per stratum, checked once per outer replicate on the observed data. Replicates violating this prespecified minimum-count requirement were discarded and the frequency and location of these violations were recorded as diagnostics of increasingly limited treatment overlap.

Overlap was varied independently of the original $X_1$-based treatment-assignment structure by adding a second independent covariate $X_2\sim N(0,1)$ to the assignment model,
\begin{equation}
\operatorname{logit}\{e(X_{1i},X_{2i})\} = \beta_0+ 0.90X_{1i} + 0.25X_{1i}^2 + \gamma X_{2i},
\end{equation}
with $\gamma\in\{0.5,1.5,2.0\}$, representing progressively poorer overlap, and $\beta_0$ again chosen to maintain treatment prevalence near 0.50. $X_2$ affected treatment assignment but not the outcome-generating process, allowing weight variability and positivity to deteriorate without strengthening the original $X_1$-outcome confounding pathway. Diagnostics included the proportions of fitted propensity scores below .05 and above .95, the observed and resampled Kish ratios, maximum and upper-tail weights, and $\Delta r_e$.

Marginal treatment prevalence was separately varied over approximately $P(A=1)\in\{.25,.50,.75\}$ at $\lambda=1$, with fixed-weight permutation and assignment-mechanism resampling evaluated under correct and misspecified propensity-score models, assessing whether unequal treatment-group sizes alter the calibration patterns observed under approximately equal prevalence.

A definitive calibration experiment then evaluated KS, AD, and CVM on the same simulated datasets under five representative conditions: an exogenous-weight reference condition; fixed estimated weights under correct propensity-model specification at $\lambda=1$ and treatment prevalence approximately .50; weight re-estimation after unrestricted label permutation under correct propensity-model specification at $\lambda=1$ and prevalence approximately .50; correctly specified assignment-mechanism resampling at prevalence approximately .50; and misspecified assignment-mechanism resampling at prevalence approximately .25. Each condition used 2,000 Monte Carlo replications and $R=500$ resampling replicates. For AD, the Kish effective sample size was recalculated from each replicate's own weight vector.

\subsubsection{Comparative power and performance measures}

A final experiment assessed whether the comparative sensitivities observed in Simulation I persisted when treatment weights were estimated. The same three forms of distributional discrepancy---central, tail, and diffuse---were retained, but the outcome-generating mechanism was extended to preserve treatment--outcome confounding:
\begin{equation}
Y_i(0) = 50+7.5X_{1i} + 10\Phi^{-1}(p_i),
\end{equation}
\begin{equation}
Y_i(1) = Y_i(0)+\delta(p_i), \qquad p_i\sim U(0,1),
\end{equation}
so that $Y_i(1)-Y_i(0)=\delta(p_i)$, and the location and magnitude of the induced distributional discrepancy are known exactly.

Three inferential schemes were compared: exogenous supplied weights, fixed estimated weights, and correctly specified assignment-mechanism resampling. The weight re-estimation procedure and misspecified assignment-mechanism condition were not included because the calibration experiments had identified potentially substantial Type I error distortion in those settings, making power difficult to interpret independently of size.

Effect magnitudes were calibrated in preliminary simulations to avoid floor and ceiling effects within each weighting scheme and were fixed before the definitive power study. For the exogenous-weight scheme, $\Delta=3$, $5.5$, and $1$ were used for the central, tail, and diffuse discrepancies, respectively; for the two estimated-weight schemes, the corresponding values were $15$, $7.5$, and $3$. Because discrepancy magnitudes therefore differ between the exogenous and estimated-weight schemes, absolute power should not be compared across those schemes; the inferential comparison of interest is the relative performance of KS, AD, and CVM within each scheme.

Each definitive power condition used 2,000 Monte Carlo replications and $R=500$ resampling replicates. All three tests were evaluated on the same simulated dataset and resampling assignments within each replication. Empirical rejection probability at $\alpha=.05$ was the primary performance measure.

Across the calibration experiments, empirical Type I error at $\alpha=.05$ was the primary performance measure. Monte Carlo standard errors were used to quantify simulation uncertainty in the empirical rejection probabilities. Weighting diagnostics included treatment prevalence, propensity-score tails, observed and resampled Kish ratios, $\Delta r_e$, and extreme-weight behavior.

Because KS, AD, and CVM were evaluated on identical datasets and resampling assignments, their rejection indicators were paired. Pairwise differences in rejection probability were evaluated using a two-sided paired $z$-test based on the mean and standard error of the per-replication rejection differences. These comparisons supplemented the empirical rejection probabilities and their Monte Carlo uncertainty.

\section{Results}\label{sec3}

\subsection{Simulation I: Exogenously supplied weights}

\subsubsection{Type I error and comparative power}

Table~\ref{tab2} reports empirical Type I error for the three weighted tests under exogenously generated case weights with a target Kish effective-sample-size ratio of $r_e=.60$. Across sample sizes from $n=1{,}000$ to $n=4{,}000$, rejection rates remained close to the nominal .05 level: KS ranged from .043 to .055, AD from .043 to .057, and CVM from .042 to .057. Thus, under weights generated independently of group assignment, fixed-weight label permutation was approximately calibrated for all three statistics across the sample sizes examined.

Table~\ref{tab3} reports empirical power against the three distributional discrepancies. Table~\ref{tab4} provides the corresponding paired comparisons of rejection rates among the three tests for both Type I error and power; none of the paired differences in Type I error was statistically significant. For the centrally located discrepancy, KS was the most powerful test at every sample size, increasing from .449 at $n=1{,}000$ to .996 at $n=4{,}000$. AD increased from .193 to .912 and CVM from .146 to .828. The paired comparisons showed KS exceeding both AD and CVM at every sample size ($p<.001$ throughout), while AD also exceeded CVM.

For the tail-located discrepancy, the ordering changed markedly. AD power increased from .114 at $n=1{,}000$ to .914 at $n=4{,}000$, whereas KS increased only from .045 to .137 and CVM from .051 to .072. AD exceeded both alternatives at every sample size ($p<.001$). KS and CVM differed little at the smaller sample sizes; at $n=4{,}000$, KS exceeded CVM by .065 ($p<.001$).

For the diffuse discrepancy, AD and CVM both outperformed KS. Across $n=1{,}000$ to $4{,}000$, power increased from .215 to .648 for AD, .205 to .618 for CVM, and .183 to .536 for KS. AD and CVM were statistically indistinguishable at $n=1{,}000$ ($p=.140$); AD subsequently held a modest advantage over CVM at the larger sample sizes. Taken together, these results reproduce the expected qualitative distinctions among the three statistics under exogenous weighting: KS was most sensitive to the centrally concentrated discrepancy, AD was substantially more sensitive to the tail discrepancy, and AD and CVM performed similarly against the diffuse discrepancy, with both outperforming KS.

\subsubsection{Sensitivity to exogenous weight variability}\label{sec3.6}

Table~\ref{tab5} extends the $n=2{,}000$ simulations across $r_e\in\{.20,.40,.60,.80,.90\}$. The comparative power patterns were similar across the range of weight variability. KS remained most powerful against the central discrepancy, AD remained substantially most powerful against the tail discrepancy, and AD and CVM remained closely matched and more powerful than KS against the diffuse discrepancy.

Type I error remained close to nominal at $r_e=.40$ through .90. At $r_e=.20$, representing the greatest weight variability examined, rejection rates increased to .064 for KS, .067 for AD, and .064 for CVM. Thus, calibration under exogenous fixed weights was generally stable across the range studied, although modest anticonservatism appeared at the most extreme weight variability.

\subsection{Simulation II: Estimated treatment weights}

\subsubsection{Calibration across resampling procedures}

Table~\ref{tab6} summarizes the calibration experiments for estimated treatment weights. The three resampling procedures behaved differently as treatment-assignment strength increased. Fixed-weight permutation remained conservative under both propensity-model specifications. Under correct specification, Type I error increased gradually from .018 at $\lambda=.5$ to .026 at $\lambda=1$ and .032 at $\lambda=1.5$, while the observed Kish ratio declined from .931 to .605. Because the same weight vector is retained throughout fixed-weight permutation, $\Delta r_e=0$ by construction. The corresponding misspecified-model rejection rates were .016, .032, and .033.

Weight re-estimation after unrestricted label permutation behaved differently. Under correct specification, the resampled Kish ratio remained approximately .994 across all three values of $\lambda$, while the observed Kish ratio declined from .931 to .615. Consequently, $\Delta r_e$ increased from .063 to .379 and Type I error increased from .057 to .274. Under propensity-model misspecification the pattern was stronger: at $\lambda=1.5$, $\Delta r_e=.473$ and Type I error reached .463. Thus, re-estimating the propensity model after unrestricted permutation did not restore calibration; as the observed and reference weighting regimes diverged, the procedure became increasingly anticonservative.

Assignment-mechanism resampling was approximately aligned with the observed weighting regime under correct propensity-model specification: $\Delta r_e$ ranged only from $-$.002 to $-$.001 across the three assignment-strength conditions, with rejection rates of .021, .023, and .033. Under misspecification, however, the mismatch increased with assignment strength. At $\lambda=1.5$, $\Delta r_e=.126$ and Type I error increased to .178. Accordingly, explicit modeling of the assignment mechanism did not by itself protect against miscalibration when the fitted propensity model inadequately represented that mechanism.

\subsubsection{Weight-regime mismatch}

Panel B of Table~\ref{tab6} more directly examined the association between observed--reference weight-regime mismatch and calibration by varying the degree to which permutation preserved the observed treatment--covariate relationship. Under correct propensity-model specification, unrestricted permutation ($S=1$) produced $\Delta r_e=.223$ and Type I error $=.120$. Increasing the number of strata reduced the mismatch: at $S=2$, $\Delta r_e=.134$ and Type I error $=.108$; at $S=4$, the regimes were approximately aligned ($\Delta r_e=-.012$) and Type I error declined to .066; and at $S=8$, $\Delta r_e=-.082$ and the test became conservative, with Type I error $=.023$. The misspecified-model conditions showed the same directional pattern, with Type I error declining from .144 at $S=1$ to .052 at $S=8$ as $\Delta r_e$ declined from .205 to .010.

The $S=16$ conditions were not directly comparable because approximately 53\% of outer replications failed the prespecified minimum stratum-count requirement and were discarded. These failures were concentrated at the extremes of $X_1$, consistent with increasingly limited treatment overlap. The lower-stratum conditions nevertheless show a clear directional relationship: positive observed--reference mismatch was associated with anticonservatism, approximate regime alignment with improved calibration, and negative mismatch with conservatism. This experiment establishes an empirical association rather than a general proof that $\Delta r_e$ alone determines calibration.

\subsubsection{Overlap and treatment prevalence}

Panels C and D of Table~\ref{tab6cont} distinguish weight variability itself from mismatch between the observed and reference weighting regimes. In the overlap experiment, progressively poorer overlap substantially reduced the observed Kish ratio without producing weight-re-estimation-like anticonservatism. Under correctly specified fixed-weight permutation, $r_e$ declined from .717 at $\gamma=.5$ to .335 at $\gamma=2$, while Type I error increased gradually from .030 to .043. Under correctly specified assignment-mechanism resampling, $\Delta r_e$ remained close to zero throughout and Type I error increased from .018 to .045. The corresponding misspecified conditions also remained near nominal at the most severe overlap level examined. These results indicate that substantial absolute weight variability alone did not reproduce the severe anticonservatism associated with large positive observed--reference mismatch.

Treatment prevalence revealed greater sensitivity to propensity-model misspecification. Fixed-weight permutation remained conservative across prevalences of .25, .50, and .75 under both propensity-model specifications, with Type I error ranging from .005 to .034. Correctly specified assignment-mechanism resampling likewise remained conservative, with rejection rates from .024 to .029 and $\Delta r_e$ close to zero. Under misspecification, however, performance depended strongly on prevalence. At prevalence .25, $\Delta r_e=.134$ and Type I error increased to .220, whereas at prevalence .75, $\Delta r_e=-.057$ and Type I error was .048. Thus, the apparently acceptable performance of assignment-mechanism resampling in one misspecified condition did not generalize across treatment prevalences.

\subsection{Calibration across KS, AD, and CVM}\label{sec3.3}

The definitive calibration experiment examined whether the resampling patterns identified using KS generalized to all three weighted EDF statistics (Table~\ref{tab7}). Under exogenous weights, empirical Type I error was .051 for KS, .046 for AD, and .045 for CVM, consistent with nominal calibration. Fixed estimated weights were conservative for all three tests, with rejection rates of .018, .019, and .015, respectively. Correctly specified assignment-mechanism resampling was similarly conservative, with rejection rates of .021, .017, and .019.

Weight re-estimation after unrestricted label permutation was anticonservative for all three tests: Type I error was .121 for KS, .082 for AD, and .117 for CVM. Under the misspecified assignment-mechanism condition at treatment prevalence .25, rejection rates increased to .221, .152, and .188, respectively. Thus, the principal calibration patterns identified with KS were not statistic-specific: fixed-weight permutation was conservative rather than anticonservative in the estimated-weight condition examined, whereas both weight re-estimation and misspecified assignment-mechanism resampling could produce substantial Type I error inflation.

The magnitude of inflation nevertheless differed among the statistics. Under weight re-estimation, the excess Type I error above .05 was .071 for KS and .032 for AD, so AD's excess was approximately 55\% lower. Under misspecified assignment-mechanism resampling, the corresponding excesses were .171 and .102, a reduction of approximately 40\%. CVM was closer to KS than AD in both conditions. AD therefore appeared less sensitive to the reference-regime distortion in these two experiments, although the degree of attenuation differed between conditions and should not be interpreted as a general correction for resampling miscalibration.

\subsection{Comparative power under exogenous and estimated weighting}

Table~\ref{tab8} reports the definitive comparative-power results. Because discrepancy magnitudes were calibrated separately for the exogenous and estimated-weight schemes, absolute power is not compared across schemes; the relevant comparisons are among KS, AD, and CVM within each scheme.

For the central discrepancy, the exogenous-weight scheme reproduced the qualitative ordering observed in Simulation I: KS had the greatest power (.277), followed by AD (.155) and CVM (.148). The ordering changed under estimated weighting, with AD becoming the most powerful of the three tests. With fixed estimated weights, power was .640 for KS, .782 for AD, and .659 for CVM; under correctly specified assignment-mechanism resampling, the corresponding values were .537, .710, and .566. Thus, AD rather than KS was the most powerful of the three tests against the calibrated central discrepancy in both estimated-weight schemes examined.

For the tail discrepancy, AD retained its comparative advantage in every scheme. Under exogenous weighting, power was .392 for KS, .667 for AD, and .420 for CVM. With fixed estimated weights, the corresponding values were .338, .562, and .341; under correctly specified assignment-mechanism resampling they were .266, .489, and .260. The pronounced tail sensitivity of AD observed in Simulation I therefore persisted under both estimated-weight procedures.

For the diffuse discrepancy, AD and CVM again outperformed KS. Under exogenous weighting, power was .115 for KS and .131 for both AD and CVM. With fixed estimated weights, power was .543, .671, and .644, respectively; under correctly specified assignment-mechanism resampling it was .482, .590, and .540. The near-equivalence of AD and CVM under exogenous weighting therefore broadly persisted, although AD showed a somewhat larger numerical advantage under the estimated-weight schemes.

Overall, two comparative-power findings from Simulation I generalized to the estimated-weight setting: AD retained its pronounced sensitivity to tail discrepancies, and AD and CVM retained an advantage over KS for diffuse discrepancies. The central-discrepancy finding did not generalize. KS was most powerful under exogenous weighting, whereas AD was most powerful under both estimated-weight schemes examined.

\section{Applied Example}\label{sec4}

\subsection{Data and weighting model}

I illustrate the proposed distributional balance tests using the maternal-smoking data analyzed by \citet{cattaneo2010} and distributed with Stata as the \texttt{cattaneo2} dataset. The data contain 4,642 observations, of whom 864 mothers smoked during pregnancy and 3,778 did not. Following Stata's documented inverse-probability-weighting example, infant birthweight was the outcome and maternal smoking was the treatment. The treatment model was estimated by logistic regression using marital status, maternal age, first-trimester prenatal care, first birth, a quadratic term for maternal age, and interactions between maternal age and marital status and between maternal age and first-trimester prenatal care.

Inverse-probability-of-treatment weights targeting the average treatment effect were obtained from the fitted treatment probabilities using the weighting formulation defined in Section~\ref{sec:sim2dgm}. The purpose of this example is not to re-estimate or interpret the causal effect of maternal smoking on birthweight, but to illustrate how the proposed weighted EDF tests can complement conventional covariate-balance diagnostics following an established propensity-score-weighting analysis.

\subsection{Conventional and distributional balance diagnostics}

Table~\ref{tabapp1} compares conventional and distributional balance diagnostics for maternal age (\texttt{mage}), which had the largest absolute raw standardized difference among the continuous baseline covariates. Before weighting, the standardized difference in maternal age was $-.300$ and the variance ratio was .882. Following inverse-probability weighting, these values changed to .041 and 1.077, respectively. Thus, by these conventional diagnostics, weighting substantially reduced the difference in mean maternal age and brought the relative variance substantially closer to one.

The weighted EDF tests provide additional information. The weighted KS statistic was $D=.3044$, the weighted AD statistic was $2.1\times10^6$, and the weighted CVM statistic was .8465. All three permutation \textit{p}-values were .0001 based on 10,000 fixed-weight label permutations. With the finite-resampling correction in Equation~(\ref{eq:permp}), .0001 is the smallest \textit{p}-value attainable with $R=10{,}000$, indicating that none of the 10,000 permuted statistics equaled or exceeded its corresponding observed statistic.

These results do not imply that the standardized difference or variance ratio incorrectly characterized balance. Rather, the diagnostics address different features of the weighted distributions. The standardized difference summarizes the difference in means, and the variance ratio compares dispersion, whereas KS, AD, and CVM compare the empirical distributions more broadly. In this example, weighting produced close agreement in the conventional mean- and variance-based diagnostics while detectable differences remained in the overall maternal-age distributions.

\begin{figure}[htbp]
\centerline{\includegraphics[width=0.85\textwidth]{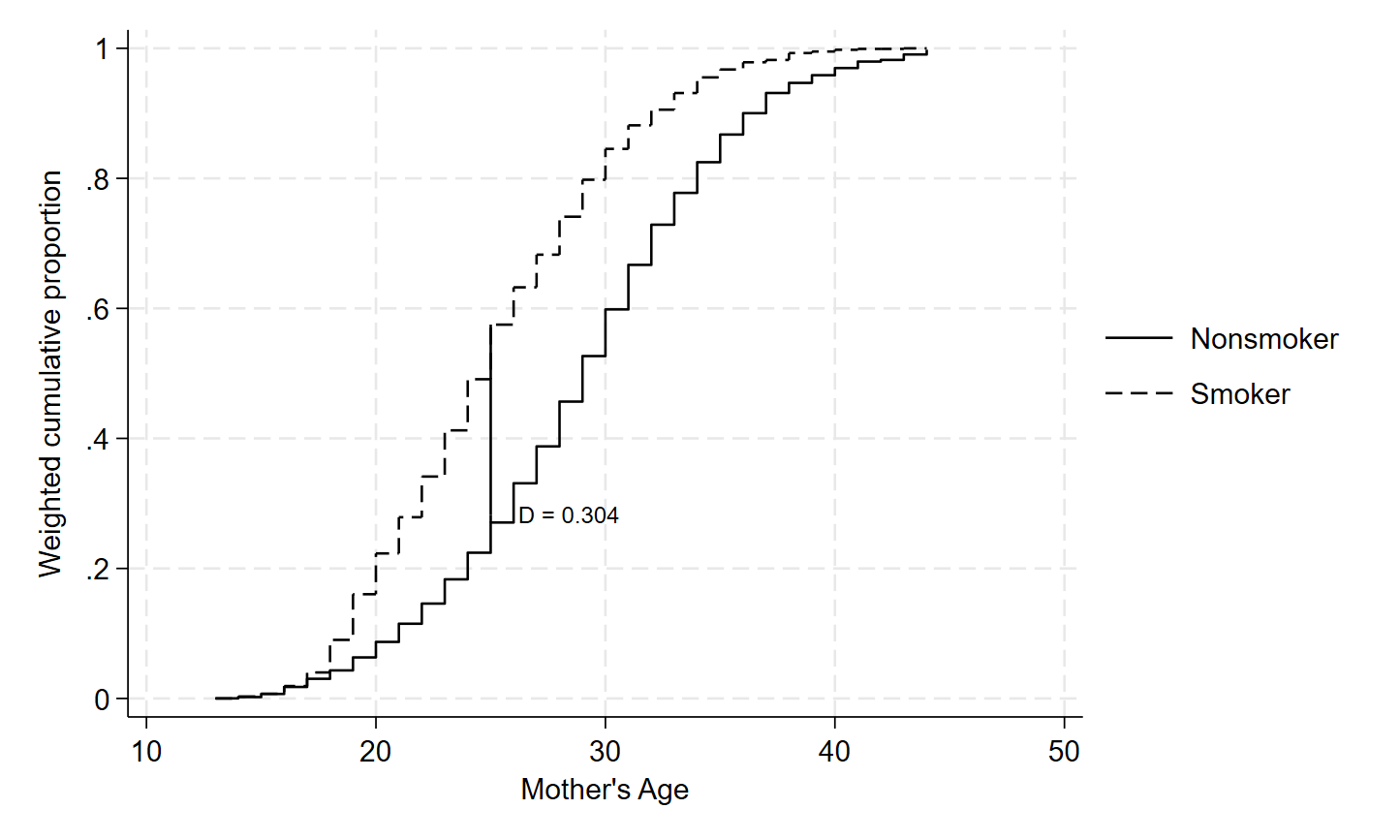}}
\caption{Weighted empirical CDFs of maternal age for smokers and nonsmokers following inverse-probability-of-treatment weighting. The location of the maximum vertical difference between the distributions, corresponding to the weighted Kolmogorov--Smirnov statistic ($D=.304$), is indicated.\label{fig1}}
\end{figure}

Figure~\ref{fig1} displays the weighted empirical CDFs of maternal age for smokers and nonsmokers. Despite the small weighted standardized difference and a variance ratio close to one, the ECDFs remain visibly separated across a substantial portion of the age distribution. The maximum vertical separation is $D=.304$, occurring in the mid-20s, but the discrepancy is not confined to that single point; separation persists across much of the central portion of the distribution before the curves converge at older ages. The graphical evidence is therefore consistent with the results of all three weighted EDF tests and illustrates the additional information obtained by examining whole-distribution balance alongside conventional moment-based diagnostics.

\section{Discussion}\label{sec5}

\subsection{Principal findings}

This study developed weighted extensions of the KS, AD, and CVM statistics and examined how their inferential behavior depends on the construction of the resampling reference distribution. Under exogenously supplied weights, fixed-weight label permutation was approximately calibrated across the sample sizes and most levels of weight variability examined. Under exogenous weighting, the comparative sensitivity patterns were also largely preserved across the conditions examined: KS was most sensitive to the centrally located discrepancy, AD was substantially more sensitive to the tail-located discrepancy, and AD and CVM outperformed KS against the diffuse discrepancy.

The estimated-weight simulations revealed a more consequential distinction. Re-estimating propensity-score weights after unrestricted label permutation could produce a reference distribution under a substantially different weighting regime from that of the observed statistic, with increasing mismatch accompanied by increasing Type I error. The restricted-permutation experiment strengthened this interpretation: calibration improved as the reference effective-information regime approached the observed regime and became conservative after the direction of the mismatch reversed. By contrast, poorer overlap and greater absolute weight variability did not themselves produce comparable anticonservatism when the observed and reference weighting regimes remained aligned.

Fixed-weight permutation behaved differently. Because the observed weights were retained throughout resampling, it avoided this source of weight-regime mismatch and was generally conservative with estimated propensity-score weights across the conditions studied. Assignment-mechanism resampling was also generally conservative when the propensity model was correctly specified, but could become anticonservative under misspecification. Thus, neither re-estimating weights nor explicitly modeling treatment assignment provides an automatic solution to reference-distribution construction when weights depend on treatment.

Finally, comparative sensitivity was not invariant to the weighting scheme. AD retained its pronounced advantage for tail discrepancies, and AD and CVM retained an advantage over KS for diffuse discrepancies. For the central discrepancy, however, KS was most powerful under exogenous weighting, whereas AD was most powerful under both estimated-weight schemes examined. Classical expectations about the relative sensitivity of these tests therefore need not carry over unchanged to treatment-dependent weighting.

Taken together, these results separate two issues that are easily conflated. The weighted EDF statistics themselves can be calculated for any supplied set of positive case weights. The inferential properties of a resampling \textit{p}-value, however, depend additionally on the relationship between those weights, treatment assignment, and the procedure used to construct the reference distribution.

\subsection{Interpretation and implications}

The distinction between defining a weighted statistic and constructing its reference distribution is especially important for propensity-score weights. Observational treatment assignments are not generally uniformly exchangeable, and previous work has emphasized conditioning on or representing the treatment-assignment mechanism in randomization-based inference \citep{rosenbaum1984,bransonbind2019,pimentelhuang2024}. The present simulations extend this issue to weighted EDF balance tests by showing that seemingly reasonable resampling choices can generate substantially different finite-sample behavior.

The weight-re-estimation results are particularly informative. Unrestricted permutation breaks the systematic treatment--covariate relationship before the propensity model is refitted. In the simulations, the resulting weights became nearly uniform even when the observed weights were substantially more variable. The direct manipulation of this mismatch through restricted permutation was accompanied by corresponding changes in Type I error. Although $\Delta r_e$ should not be interpreted as a sufficient condition for calibration, these results suggest that preserving relevant features of the observed weighting regime is important when constructing the reference distribution.

The applied example illustrates the complementary role of distributional diagnostics. In the Cattaneo maternal-smoking data, weighting reduced the maternal-age standardized difference from $-.300$ to .041 and moved the variance ratio from .882 to 1.077, indicating substantial improvement according to conventional mean- and variance-based diagnostics. Nevertheless, all three EDF tests produced $p=.0001$, and the weighted ECDFs showed visible residual separation. These findings are not contradictory: standardized differences, variance ratios, and EDF comparisons characterize different features of the weighted distributions.

The proposed tests should therefore be used as complementary diagnostics rather than pass--fail criteria for balance. Neither a small \textit{p}-value alone establishes substantively important imbalance nor a large \textit{p}-value establishes causal validity. SMDs, variance ratios, overlap and weight diagnostics, effective sample size, and graphical inspection remain important components of balance assessment. For continuous and ordered covariates, weighted EDF statistics add a whole-distribution perspective that may reveal differences not represented by conventional moment-based measures.

The results also argue against a rigid rule for choosing among KS, AD, and CVM. AD showed the most consistent performance across the estimated-weight simulations and retained its strong tail sensitivity, but no statistic was uniformly most powerful. When feasible, considering more than one EDF measure together with the weighted ECDF is more informative than choosing a test solely from its classical sensitivity profile. When a particular distributional test is selected, investigators should report the basis for that choice, recognizing that KS, AD, and CVM emphasize different features of the distribution and therefore may differ in their sensitivity to particular forms of residual imbalance.

\subsection{Limitations}

The principal limitation is that the inferential findings are simulation-based rather than a formal proof. Although Simulation II varied assignment strength, propensity-model specification, overlap, weight variability, and treatment prevalence, other propensity estimators, weighting estimands, covariate distributions, and treatment-assignment mechanisms could produce different behavior. Likewise, $\Delta r_e$ is a useful summary of the observed--reference mismatch studied here but cannot characterize every relevant feature of two weight distributions.

The power experiments deliberately used stylized central, tail, and diffuse discrepancies to isolate differences among the tests, and effect magnitudes were calibrated separately in the exogenous- and estimated-weight schemes. Absolute power therefore should not be compared across those schemes. Finally, the proposed tests assess covariates marginally; as with other univariate balance diagnostics, they neither establish joint covariate balance nor address unmeasured confounding.

\subsection{Conclusion}

Weighted KS, AD, and CVM statistics provide complementary whole-distribution diagnostics that can identify residual covariate differences not captured by conventional mean- and variance-based measures. The simulations also show that when weights depend on treatment assignment, construction of the statistic and calibration of its reference distribution are distinct problems: fixed-weight permutation was generally conservative in the estimated-weight settings studied, whereas alternative resampling procedures could become anticonservative when the reference weighting regime or assignment model was poorly aligned with the observed data. Thus, among the resampling approaches examined, fixed-weight permutation appears to be the more reliable choice for estimated propensity-score weights, with the qualification being that its conservative behavior may reduce sensitivity. No single EDF test was uniformly most sensitive across weighting schemes and discrepancy types. Investigators should therefore use these tests alongside conventional balance diagnostics and graphical assessment, and select and report distributional tests according to the features of imbalance they are intended to detect rather than treating any single test or \textit{p}-value as a universal criterion for covariate balance.

\section*{Author contributions}
Ariel Linden is the sole author of this work and is responsible for its conception, methodological development, software implementation, simulation design, analysis, and writing.

\section*{Acknowledgments}
None reported.

\section*{Financial disclosure}
None reported.

\section*{Conflict of interest}
The author declares no potential conflict of interest.

\section*{Data availability statement}
The Stata commands \texttt{kstest}, \texttt{adtest}, and \texttt{cvmtest} implementing the methods described in this paper are freely available for download within Stata. All simulation code used in this paper is available at \url{https://github.com/ariellinden/weighted_distribution_tests}.

\bibliographystyle{apalike}
\bibliography{refs}

\clearpage

\begin{table}[htbp]
\centering
\caption{Simulation Study Design\label{tab1}}
\small
\begin{tabular}{@{}p{.20\textwidth}p{.41\textwidth}p{.33\textwidth}@{}}
\toprule
Simulation component & Design conditions & Purpose \\
\midrule
\multicolumn{3}{l}{\textit{Panel A: Simulation I---Exogenously supplied weights}} \\
Type I error & $\delta(p)\equiv0$ & Assess null calibration \\
Central discrepancy & $p_L=.25,\ p_U=.75,\ w=.05,\ \Delta=6$ & Compare sensitivity to centrally concentrated imbalance \\
Tail discrepancy & $p_L=.05,\ p_U=.95,\ w=.02,\ \Delta=50$ & Compare sensitivity to tail imbalance \\
Diffuse discrepancy & $\delta(p)\equiv1$ & Compare sensitivity to a uniform location shift \\
Weight variability & $r_e=.20,.40,.60,.80,.90$ & Assess sensitivity to exogenous weight variability \\
\addlinespace
\multicolumn{3}{l}{\textit{Panel B: Simulation II---Estimated treatment weights}} \\
Assignment strength & $\lambda=.5,1,1.5$; correct and misspecified propensity models & Assess calibration as treatment--covariate dependence increases \\
Weight-regime mismatch & $S=1,2,4,8,16$ $X_1$-quantile strata; $\lambda=1$ & Assess calibration as the reference weighting regime approaches the observed regime \\
Overlap and weight variability & $\gamma=.5,1.5,2$; correct and misspecified propensity models & Separate effects of absolute weight variability from observed--reference mismatch \\
Treatment prevalence & $P(A=1)\approx.25,.50,.75$; correct and misspecified propensity models & Assess calibration with unequal treatment-group prevalence \\
Definitive calibration & Exogenous weights; fixed estimated weights; weight re-estimation; assignment-mechanism resampling (correct and misspecified) & Compare Type I error across KS, AD, and CVM \\
Comparative power & Central, tail, and diffuse discrepancies; exogenous weights, fixed estimated weights, and correctly specified assignment-mechanism resampling & Compare relative sensitivity of KS, AD, and CVM \\
\bottomrule
\end{tabular}
\begin{tablenotes}
\item {\it Note}: In Simulation I, total sample size was $n\in\{1{,}000,2{,}000,3{,}000,4{,}000\}$, with $n_1/n=.4$ and $n_0/n=.6$; the principal simulations used $r_e=.60$, 2,000 Monte Carlo replications, and $R=1{,}000$ fixed-weight permutations. In Simulation II, unless otherwise stated, $n=500$, treatment prevalence was approximately .50, and $R=500$ resampling replicates were used. Calibration experiments used 1,000 Monte Carlo replications; definitive calibration and power experiments used 2,000. Estimated weights were inverse-probability-of-treatment weights targeting the average treatment effect. See Section~\ref{sec2.6} for complete data-generating mechanisms and parameter definitions.
\end{tablenotes}
\end{table}

\clearpage

\begin{table}[htbp]
\centering
\caption{Empirical Type I Error Rates ($\alpha=0.05$)\label{tab2}}
\begin{tabular*}{\textwidth}{@{\extracolsep\fill}lccc@{}}
\toprule
$n$ & KS & AD & CVM \\
\midrule
1000 & .055 (.045, .065) & .048 (.039, .057) & .050 (.040, .060) \\
2000 & .044 (.035, .053) & .044 (.035, .053) & .046 (.037, .055) \\
3000 & .043 (.034, .052) & .043 (.034, .052) & .042 (.033, .051) \\
4000 & .055 (.045, .065) & .057 (.047, .067) & .057 (.047, .067) \\
\bottomrule
\end{tabular*}
\begin{tablenotes}
\item {\it Note}: 2,000 replications per cell; values in parentheses are 95\% confidence intervals ($\hat p \pm 1.96\sqrt{\hat p(1-\hat p)/2000}$).
\end{tablenotes}
\end{table}

\clearpage

\begin{table}[htbp]
\centering
\caption{Empirical Power Against Three Types of Distributional Discrepancy\label{tab3}}
\begin{tabular*}{\textwidth}{@{\extracolsep\fill}llccc@{}}
\toprule
Scenario & $n$ & KS & AD & CVM \\
\midrule
Centrally located & 1000 & .449 (.427, .471) & .193 (.176, .210) & .146 (.131, .161) \\
                  & 2000 & .840 (.824, .856) & .448 (.426, .470) & .323 (.303, .343) \\
                  & 3000 & .973 (.966, .980) & .703 (.683, .723) & .579 (.557, .601) \\
                  & 4000 & .996 (.993, .999) & .912 (.900, .924) & .828 (.811, .845) \\
\addlinespace
Tail-located      & 1000 & .045 (.036, .054) & .114 (.100, .128) & .051 (.041, .061) \\
                  & 2000 & .049 (.040, .058) & .297 (.277, .317) & .060 (.050, .070) \\
                  & 3000 & .081 (.069, .093) & .655 (.634, .676) & .072 (.061, .083) \\
                  & 4000 & .137 (.122, .152) & .914 (.902, .926) & .072 (.061, .083) \\
\addlinespace
Diffuse           & 1000 & .183 (.166, .200) & .215 (.197, .233) & .205 (.187, .223) \\
                  & 2000 & .309 (.289, .329) & .376 (.355, .397) & .358 (.337, .379) \\
                  & 3000 & .441 (.419, .463) & .512 (.490, .534) & .495 (.473, .517) \\
                  & 4000 & .536 (.514, .558) & .648 (.627, .669) & .618 (.597, .639) \\
\bottomrule
\end{tabular*}
\begin{tablenotes}
\item {\it Note}: 2,000 replications per cell; rejection rate at $\alpha=0.05$ using each test's permutation \textit{p}-value ($R=1{,}000$ replicates). Values in parentheses are 95\% confidence intervals ($\hat p \pm 1.96\sqrt{\hat p(1-\hat p)/2000}$). See Table~\ref{tab4} for paired-difference significance tests.
\end{tablenotes}
\end{table}

\clearpage

\begin{table}[htbp]
\centering
\caption{Paired Differences in Rejection Rate Between Tests\label{tab4}}
\small
\begin{tabular*}{\textwidth}{@{\extracolsep\fill}llrlrlrl@{}}
\toprule
& & \multicolumn{2}{c}{AD $-$ KS} & \multicolumn{2}{c}{CVM $-$ KS} & \multicolumn{2}{c}{AD $-$ CVM} \\
\cmidrule(lr){3-4}\cmidrule(lr){5-6}\cmidrule(lr){7-8}
Scenario & $n$ & $\Delta$ & $p$ & $\Delta$ & $p$ & $\Delta$ & $p$ \\
\midrule
Type I error & 1000 & $-$.007 & .262 & $-$.005 & .369 & $-$.002 & .593 \\
             & 2000 & .000 & 1.000 & .002 & .670 & $-$.002 & .617 \\
             & 3000 & .000 & 1.000 & $-$.001 & .847 & .001 & .782 \\
             & 4000 & .002 & .746 & .002 & .715 & .000 & 1.000 \\
\addlinespace
Centrally located & 1000 & $-$.256 & $<$.001 & $-$.303 & $<$.001 & .047 & $<$.001 \\
                  & 2000 & $-$.392 & $<$.001 & $-$.517 & $<$.001 & .125 & $<$.001 \\
                  & 3000 & $-$.270 & $<$.001 & $-$.394 & $<$.001 & .124 & $<$.001 \\
                  & 4000 & $-$.084 & $<$.001 & $-$.168 & $<$.001 & .084 & $<$.001 \\
\addlinespace
Tail-located & 1000 & .069 & $<$.001 & .006 & .273 & .063 & $<$.001 \\
             & 2000 & .248 & $<$.001 & .011 & .093 & .237 & $<$.001 \\
             & 3000 & .574 & $<$.001 & $-$.009 & .225 & .583 & $<$.001 \\
             & 4000 & .777 & $<$.001 & $-$.065 & $<$.001 & .842 & $<$.001 \\
\addlinespace
Diffuse & 1000 & .032 & .001 & .022 & .003 & .010 & .140 \\
        & 2000 & .067 & $<$.001 & .049 & $<$.001 & .018 & .016 \\
        & 3000 & .071 & $<$.001 & .054 & $<$.001 & .017 & .029 \\
        & 4000 & .112 & $<$.001 & .082 & $<$.001 & .030 & $<$.001 \\
\bottomrule
\end{tabular*}
\begin{tablenotes}
\item {\it Note}: $\Delta$ is the mean per-replicate difference in rejection outcome (e.g., AD$-$KS: positive values indicate AD rejected more often than KS on the same simulated datasets); $p$-values are from a paired $z$-test using the standard deviation of the per-replicate difference, which correctly accounts for the correlation between tests run on identical data. $p<.001$ shown where the exact value rounds to $.000$ at three decimal places.
\end{tablenotes}
\end{table}

\clearpage

\begin{table}[htbp]
\centering
\caption{Weight-Variability Sensitivity: Empirical Type I Error and Power by Condition and $r_e$ ($n=2{,}000$)\label{tab5}}
\begin{tabular*}{\textwidth}{@{\extracolsep\fill}llccc@{}}
\toprule
Condition & $r_e$ & KS & AD & CVM \\
\midrule
Type I error (null)         & 0.2 & .064 & .067 & .064 \\
                             & 0.4 & .048 & .064 & .056 \\
                             & 0.6 & .044 & .044 & .046 \\
                             & 0.8 & .046 & .041 & .041 \\
                             & 0.9 & .056 & .054 & .061 \\
\addlinespace
Power: centrally located    & 0.2 & .317 & .138 & .108 \\
                             & 0.4 & .576 & .251 & .193 \\
                             & 0.6 & .840 & .448 & .323 \\
                             & 0.8 & .953 & .623 & .497 \\
                             & 0.9 & .978 & .694 & .571 \\
\addlinespace
Power: tail-located          & 0.2 & .052 & .101 & .059 \\
                             & 0.4 & .053 & .182 & .068 \\
                             & 0.6 & .049 & .297 & .060 \\
                             & 0.8 & .064 & .538 & .066 \\
                             & 0.9 & .065 & .653 & .063 \\
\addlinespace
Power: diffuse               & 0.2 & .144 & .164 & .161 \\
                             & 0.4 & .206 & .252 & .239 \\
                             & 0.6 & .309 & .376 & .358 \\
                             & 0.8 & .379 & .433 & .438 \\
                             & 0.9 & .439 & .526 & .513 \\
\bottomrule
\end{tabular*}
\begin{tablenotes}
\item {\it Note}: 2,000 replications per cell; sample size fixed at $n=2{,}000$ ($n_1=800$, $n_0=1200$) throughout, extending the single value of $r_e=0.6$ used in Table~\ref{tab2} and Table~\ref{tab3} to a range from milder ($r_e=0.9$, $0.8$) to more severe ($r_e=0.4$, $0.2$) weight variability. All values are rejection rates at $\alpha=0.05$: for the ``Type I error'' rows, this is the empirical size under the null (no true discrepancy); for the three ``Power'' rows, this is empirical power against the discrepancy named. The $r_e=0.6$ row in each panel reproduces the corresponding $n=2{,}000$ cell from Table~\ref{tab2} (Type I error) or Table~\ref{tab3} (power).
\end{tablenotes}
\end{table}

\clearpage

\begin{table}[htbp]
\centering
\caption{Simulation II Calibration Experiments: Assignment Strength and Weight-Regime Mismatch\label{tab6}}
\footnotesize
\begin{tabular*}{\textwidth}{@{\extracolsep\fill}lccccc@{}}
\toprule
Condition & Spec. & Observed $r_e$ & Resampled $r_e$ & $\Delta r_e$ & Type I error \\
\midrule
\multicolumn{6}{l}{\textit{Panel A: Assignment strength ($\lambda$)}} \\
\multicolumn{6}{l}{\hspace{1em}\textit{Fixed-weight permutation}} \\
$\lambda=0.5$  & Correct   & .931 & .931 & 0     & .018 \\
$\lambda=1.0$  & Correct   & .772 & .772 & 0     & .026 \\
$\lambda=1.5$  & Correct   & .605 & .605 & 0     & .032 \\
$\lambda=0.5$  & Misspec.  & .947 & .947 & 0     & .016 \\
$\lambda=1.0$  & Misspec.  & .789 & .789 & 0     & .032 \\
$\lambda=1.5$  & Misspec.  & .529 & .529 & 0     & .033 \\
\addlinespace[2pt]
\multicolumn{6}{l}{\hspace{1em}\textit{Weight re-estimation after label permutation}} \\
$\lambda=0.5$  & Correct   & .931 & .994 & .063 & .057 \\
$\lambda=1.0$  & Correct   & .772 & .994 & .222 & .122 \\
$\lambda=1.5$  & Correct   & .615 & .994 & .379 & .274 \\
$\lambda=0.5$  & Misspec.  & .945 & .996 & .051 & .066 \\
$\lambda=1.0$  & Misspec.  & .791 & .996 & .205 & .158 \\
$\lambda=1.5$  & Misspec.  & .523 & .996 & .473 & .463 \\
\addlinespace[2pt]
\multicolumn{6}{l}{\hspace{1em}\textit{Assignment-mechanism resampling}} \\
$\lambda=0.5$  & Correct   & .932 & .930 & $-$.002 & .021 \\
$\lambda=1.0$  & Correct   & .765 & .764 & $-$.001 & .023 \\
$\lambda=1.5$  & Correct   & .608 & .607 & $-$.002 & .033 \\
$\lambda=0.5$  & Misspec.  & .946 & .947 & .001  & .023 \\
$\lambda=1.0$  & Misspec.  & .794 & .823 & .030  & .051 \\
$\lambda=1.5$  & Misspec.  & .531 & .658 & .126  & .178 \\
\addlinespace[2pt]
\multicolumn{6}{l}{\textit{Panel B: Observed--reference mismatch via restricted permutation ($\lambda=1$)}} \\
$S=1$  & Correct  & .770 & .994 & .223 & .120 \\
$S=2$  & Correct  & .762 & .896 & .134 & .108 \\
$S=4$  & Correct  & .771 & .758 & $-$.012 & .066 \\
$S=8$  & Correct  & .767 & .686 & $-$.082 & .023\rlap{$^{a}$} \\
$S=16$ & Correct  & .782 & .715 & $-$.067 & .019\rlap{$^{a}$} \\
$S=1$  & Misspec. & .791 & .996 & .205 & .144 \\
$S=2$  & Misspec. & .794 & .913 & .119 & .137 \\
$S=4$  & Misspec. & .788 & .826 & .037 & .096 \\
$S=8$  & Misspec. & .791 & .800 & .010 & .052\rlap{$^{a}$} \\
$S=16$ & Misspec. & .806 & .808 & .002 & .047\rlap{$^{a,b}$} \\
\bottomrule
\end{tabular*}
\begin{tablenotes}
\item {\it Note}: 1,000 Monte Carlo replications and $R=500$ resampling replicates per condition ($n=500$ throughout), except Panel B, which used $R=500$ resamples but is summarized here from a design in which observed and reference Kish ratios were computed only for the retained (non-discarded) outer replications. For fixed-weight permutation, the resampled weight vector is identical to the observed vector by construction; consequently $r_{e,\mathrm{resamp}}=r_{e,\mathrm{obs}}$ and $\Delta r_e=0$. Panels C and D (overlap and treatment prevalence) appear in the following table.
\item $^{a}$ At $S=8$ and $S=16$, some outer replications were discarded for violating the minimum stratum-count requirement (at least 3 treated and 3 control per stratum); retained samples were 994--995/1{,}000 at $S=8$ and 468--472/1{,}000 at $S=16$.
\item $^{b}$ Given the substantial discard rate at $S=16$ (approximately 53\%), concentrated at the extremes of the $X_1$ distribution, this condition is best read as a selected-sample diagnostic rather than as directly comparable to the lower-stratum conditions.
\end{tablenotes}
\end{table}

\clearpage

\begin{table}[htbp]
\centering
\caption{Simulation II Calibration Experiments: Overlap and Treatment Prevalence\label{tab6cont}}
\footnotesize
\begin{tabular*}{\textwidth}{@{\extracolsep\fill}lccccc@{}}
\toprule
Condition & Spec. & Observed $r_e$ & Resampled $r_e$ & $\Delta r_e$ & Type I error \\
\midrule
\multicolumn{6}{l}{\textit{Panel C: Overlap and weight variability ($\gamma$)}} \\
\multicolumn{6}{l}{\hspace{1em}\textit{Fixed-weight permutation}} \\
$\gamma=0.5$ & Correct  & .717 & .717 & 0     & .030 \\
$\gamma=1.5$ & Correct  & .442 & .442 & 0     & .036 \\
$\gamma=2.0$ & Correct  & .335 & .335 & 0     & .043 \\
$\gamma=0.5$ & Misspec. & .733 & .733 & 0     & .029 \\
$\gamma=1.5$ & Misspec. & .451 & .451 & 0     & .030 \\
$\gamma=2.0$ & Misspec. & .348 & .348 & 0     & .044 \\
\addlinespace[2pt]
\multicolumn{6}{l}{\hspace{1em}\textit{Assignment-mechanism resampling}} \\
$\gamma=0.5$ & Correct  & .727 & .717 & $-$.011 & .018 \\
$\gamma=1.5$ & Correct  & .431 & .445 & .014  & .037 \\
$\gamma=2.0$ & Correct  & .331 & .338 & .007  & .045 \\
$\gamma=0.5$ & Misspec. & .735 & .762 & .027  & .044 \\
$\gamma=1.5$ & Misspec. & .454 & .463 & .009  & .035 \\
$\gamma=2.0$ & Misspec. & .338 & .350 & .012  & .052 \\
\addlinespace[2pt]
\multicolumn{6}{l}{\textit{Panel D: Treatment prevalence}} \\
\multicolumn{6}{l}{\hspace{1em}\textit{Fixed-weight permutation}} \\
Prev.\ = .25 & Correct  & .529 & .529 & 0     & .005 \\
Prev.\ = .50 & Correct  & .771 & .771 & 0     & .020 \\
Prev.\ = .75 & Correct  & .509 & .509 & 0     & .018 \\
Prev.\ = .25 & Misspec. & .350 & .350 & 0     & .012 \\
Prev.\ = .50 & Misspec. & .793 & .793 & 0     & .024 \\
Prev.\ = .75 & Misspec. & .653 & .653 & 0     & .034 \\
\addlinespace[2pt]
\multicolumn{6}{l}{\hspace{1em}\textit{Assignment-mechanism resampling}} \\
Prev.\ = .25 & Correct  & .528 & .530 & .002  & .024 \\
Prev.\ = .50 & Correct  & .774 & .768 & $-$.006 & .029 \\
Prev.\ = .75 & Correct  & .501 & .508 & .007  & .028 \\
Prev.\ = .25 & Misspec. & .351 & .485 & .134  & .220 \\
Prev.\ = .50 & Misspec. & .791 & .821 & .031  & .035 \\
Prev.\ = .75 & Misspec. & .652 & .596 & $-$.057 & .048 \\
\bottomrule
\end{tabular*}
\begin{tablenotes}
\item {\it Note}: 1,000 Monte Carlo replications and $R=500$ resampling replicates per condition ($n=500$ throughout). For fixed-weight permutation, the resampled weight vector is identical to the observed vector by construction; consequently $r_{e,\mathrm{resamp}}=r_{e,\mathrm{obs}}$ and $\Delta r_e=0$, as in the preceding table. Panels A and B (assignment strength and weight-regime mismatch) appear in the preceding table.
\end{tablenotes}
\end{table}

\clearpage

\begin{table}[htbp]
\centering
\caption{Simulation II: Definitive Type I Error Across KS, AD, and CVM\label{tab7}}
\begin{tabular*}{\textwidth}{@{\extracolsep\fill}lccccc@{}}
\toprule
Condition & KS & AD & CVM & Observed $r_e$ & $\Delta r_e$ \\
\midrule
Exogenous reference                    & .051 & .046 & .045 & .605 & 0 \\
Fixed estimated weights                & .018 & .019 & .015 & .772 & 0 \\
Weight re-estimation                   & .121 & .082 & .117 & .768 & .226 \\
Assignment mechanism, correct          & .021 & .017 & .019 & .769 & $-$.005 \\
Assignment mechanism, misspecified     & .221 & .152 & .188 & .344 & .140 \\
\bottomrule
\end{tabular*}
\begin{tablenotes}
\item {\it Note}: 2,000 Monte Carlo replications and $R=500$ resampling replicates per condition ($n=500$; fixed estimated weights and weight re-estimation used correctly specified propensity models at $\lambda=1$ and prevalence approximately .50; assignment-mechanism conditions used prevalence approximately .50 [correct] and .25 [misspecified], per Section~\ref{sec2.6}). For fixed-weight permutation, $r_{e,\mathrm{resamp}}=r_{e,\mathrm{obs}}$ and $\Delta r_e=0$ by construction.
\end{tablenotes}
\end{table}

\clearpage

\begin{table}[htbp]
\centering
\caption{Simulation II: Definitive Comparative Power Across KS, AD, and CVM\label{tab8}}
\begin{tabular*}{\textwidth}{@{\extracolsep\fill}llcccc@{}}
\toprule
Discrepancy & Procedure & $\Delta$ & KS & AD & CVM \\
\midrule
Central & Exogenous weights              & 3  & .277 & .155 & .148 \\
        & Fixed estimated weights        & 15 & .640 & .782 & .659 \\
        & Assignment mechanism resampling  & 15 & .537 & .710 & .566 \\
\addlinespace
Tail    & Exogenous weights              & 5.5 & .392 & .667 & .420 \\
        & Fixed estimated weights        & 7.5 & .338 & .562 & .341 \\
        & Assignment mechanism resampling  & 7.5 & .266 & .489 & .260 \\
\addlinespace
Diffuse & Exogenous weights              & 1  & .115 & .131 & .131 \\
        & Fixed estimated weights        & 3  & .543 & .671 & .644 \\
        & Assignment mechanism resampling  & 3  & .482 & .590 & .540 \\
\bottomrule
\end{tabular*}
\begin{tablenotes}
\item {\it Note}: 2,000 Monte Carlo replications and $R=500$ resampling replicates per condition ($n=500$ throughout). Values are empirical rejection probability at $\alpha=.05$. Discrepancy magnitudes ($\Delta$) were calibrated separately within the exogenous and estimated-weight schemes to avoid floor and ceiling effects (Section~\ref{sec2.6}) and therefore differ between them; consequently, absolute power should not be compared across schemes, only KS, AD, and CVM within a given scheme. The two estimated-weight schemes use identical $\Delta$ within each discrepancy type.
\end{tablenotes}
\end{table}

\clearpage

\begin{table}[htbp]
\centering
\caption{Balance Diagnostics for Maternal Age Before and After Inverse-Probability Weighting\label{tabapp1}}
\begin{tabular*}{\textwidth}{@{\extracolsep\fill}lcc@{}}
\toprule
Diagnostic & Raw & Weighted \\
\midrule
Standardized difference & $-$.300 & .041 \\
Variance ratio           & .882    & 1.077 \\
\addlinespace
\multicolumn{3}{l}{\textit{Weighted distributional tests}} \\
 & Statistic & Permutation $p$ \\
KS  & $D = .3044$          & .0001 \\
AD  & $AD = 2.1\times10^6$ & .0001 \\
CVM & $CVM = .8465$        & .0001 \\
\bottomrule
\end{tabular*}
\begin{tablenotes}
\item {\it Note}: Standardized differences and variance ratios are from \texttt{tebalance summarize} following inverse-probability weighting. Distributional tests use the estimated inverse-probability weights and $R=10{,}000$ fixed-weight label permutations. With the finite-resampling correction used here, $p=.0001$ is the minimum attainable \textit{p}-value. The diagnostics characterize different aspects of covariate balance and should not be interpreted as interchangeable criteria for declaring balance or imbalance.
\end{tablenotes}
\end{table}

\end{document}